\documentclass[11pt]{evolution}    

\usepackage{natbib}

\usepackage{epsfig,amsmath,amsfonts,amssymb,color} 
\definecolor{red}{rgb}{1.0,0.0,0.0}

\newcommand{\be}{\begin{equation}}
\newcommand{\ee}{\end{equation}}
\newcommand{\bea}{\begin{eqnarray}}
\newcommand{\eea}{\end{eqnarray}}
\newcommand{\bml}{\begin{mathletters}}
\newcommand{\eml}{\end{mathletters}}

\renewcommand{\citep}[1]{(\citealt{#1})}

\raggedright

\begin{document}

\title{Fixation of mutators in asexual
  populations: the role of genetic drift and epistasis} 

\author{Kavita Jain$^{1,*}$%
       \email{Kavita Jain - jain@jncasr.ac.in}%
      and
         Apoorva Nagar$^2$%
         \email{Apoorva Nagar - madappu@hotmail.com}
}

\address{%
    \iid(1)Theoretical Sciences Unit and  Evolutionary and
Organismal Biology Unit,
Jawaharlal Nehru Centre for Advanced Scientific Research, Jakkur P.O.,
Bangalore 560064, India\\
    \iid(2) Department of Physics,
Indian Institute of  Space Science and Technology,
Valiamala P.O.,
Thiruvananthapuram, Kerala, India
}%

\maketitle
Running Title : Fixation of mutators

\vspace{5mm}
{\bf Contact Information (for all authors)}

Kavita Jain 

{\bf postal address}: Theoretical Sciences Unit and  Evolutionary and
Organismal Biology Unit,
Jawaharlal Nehru Centre for Advanced Scientific Research, Jakkur P.O.,
Bangalore 560064, India

{\bf work telephone number}: +91-80-22082948

{\bf E-mail}:jain@jncasr.ac.in

Apoorva Nagar

{\bf postal address}: Department of Physics,
Indian Institute of  Space Science and Technology,
Valiamala P.O.,
Thiruvananthapuram, Kerala, India

{\bf work telephone number}: +91-471-2568462

{\bf E-mail}: madappu@hotmail.com

\clearpage
\begin{abstract}
We study the evolutionary dynamics of  an asexual
population of nonmutators and mutators on a class of epistatic fitness
landscapes. We consider the 
situation in which all mutations are deleterious and mutators are
produced from nonmutators continually at a constant rate.  
We find that in an infinitely large population, a minimum
nonmutator-to-mutator conversion rate is required to fix the 
mutators but an arbitrarily small conversion rate results in the
fixation of mutators in a 
finite population.   
We calculate analytical expressions for the mutator 
fraction at mutation-selection balance and fixation time for mutators
in a finite population 
when the  difference between the mutation rate for mutator
  and nonmutator 
  is smaller (regime I) and larger (regime II) than the selection
  coefficient.  
%al effects are weaker (regime I) and stronger
%(regime II) than the selective effects. 
Our main result is that in
regime I, the
mutator fraction and the fixation time are independent of epistasis
but in regime II,  mutators are rarer and take longer to fix when the
decrease in fitness with the number of deleterious mutations occurs at an
accelerating rate (synergistic epistasis) than at a 
diminishing rate  (antagonistic epistasis). Our
analytical results are compared with numerics and their  implications
are discussed.   

\vspace{1cm}
\end{abstract}
\clearpage

%============================================================================
%INTRODUCTION
%============================================================================

Mutators are those cells or genotypes that have a higher mutation rate
than the wild type \citep{Visser:2002,Baer:2007}. While transient mutators
arise and decline in response to an external perturbation such  
as stress, constitutive mutators carry defect(s) in
proofreading and  mismatch repair pathways and are heritable
\citep{Miller:1996,Rosenberg:1998}. The 
latter class of mutators which we consider here have been seen 
in several microbial populations in natural 
\citep{LeClerc:1996,Matic:1997,Oliver:2000,Bjorkholm:2001,Richardson:2002} 
and experimental \citep{Mao:1997,Sniegowski:1997,Boe:2000} settings.  
The mutator fequency is seen to vary widely from less than a percent
\citep{Gross:1981,Trobner:1984,Mao:1997}  
to hundred percent \citep{Mao:1997,Sniegowski:1997}. 

As most mutations are deleterious \citep{Drake:1998}, how do mutators
manage to reach significant frequencies? 
In a maladapted population, 
while the mutators produce more deleterious mutations per generation 
compared to the nonmutators, they also give rise to rare beneficial 
mutations more often. In an asexual population in which 
the mutator and its effects remain linked, the mutator population
can {\it hitchhike} \citep{Smith:1974} with beneficial mutations to high
frequencies and eventually get fixed 
\citep{Sniegowski:1997,Shaver:2002}. 
This situation has been a subject of many theoretical
\citep{Taddei:1997,Tenaillon:1999,Palmer:2006,Andre:2006} and experimental
studies
\citep{Trobner:1984,Sniegowski:1997,Cooper:2000,Shaver:2002,Gentile:2011}.

Here we are interested in the dynamics of an asexual population of
nonmutators and mutators {\it after} the adaptation process is over. We
assume that only deleterious mutations can occur and neglect rare 
beneficial mutations arising due to back or compensatory mutations and 
the creation of nonmutators from mutators 
\citep{Drake:1998}. In this 
setting, clearly the mutators cannot go to fixation by hitchhiking. 
However if the mutators are being continually produced 
and the population size is finite, it
is obvious that they will eventually get fixed for {\it arbitrarily}
small conversion rate. It is important to note that in an infinitely
large population at mutation-selection balance, a {\it minimum}
nonmutator-to-mutator conversion rate is required above which the
mutators get fixed \citep{Tannenbaum:2005,Nagar:2009}.

Some recent works \citep{Soderberg:2011,Lynch:2011} have studied the
dynamics of mutator fixation in  
finite populations when beneficial mutations are absent assuming that
the fitnesses do not interact epistatically and the mutational effects 
are weaker than the selective costs. 
However experiments indicate that 
epistasis is rather common 
\citep{Visser:2007,Kouyos:2007,Phillips:2008} and
theoretical studies have shown that it 
strongly affects both statics and dynamics
\citep{Jain:2008b,Jain:2010a}. To our knowledge, the effect of
epistatic interactions on mutator dynamics has not been 
explored in previous theoretical studies. On the experimental
front, strong epistatic interactions between deleterious mutations have been
observed in {\it S. typhimurium} for a broad range of mutation rates
\citep{Maisnier-Patin:2005}.

In this article, we consider the evolution of both infinitely large
and finite sized populations on a class of epistatic fitness
landscapes. The epistatic interaction between the fitnesses is
antagonistic (synergistic) if the fitness of the double
mutant is larger (smaller) than the sum of the fitnesses of the two
single mutants. Using a continuous time, deterministic model, we first 
show 
that in an infinite population, a transition occurs at a critical 
conversion rate between a mixed state with both nonmutators and
mutators and a pure mutator state for any epistasis. We then develop
 a diffusion theory to calculate the time to fix mutators when the
 conversion rate is small. We find that the fixation time depends
 weakly on the population size for small populations but grows
 exponentially fast for larger populations. The population size at
 which this crossover occurs 
 and the growth rate of fixation time for large populations are
 governed by the mutator 
 fraction at  mutation-selection balance which we analytically
 calculate when mutational effects are weaker (regime I) and
 stronger (regime II) than the selective effects. We find that in regime
 I, the  mutator fraction is independent of the epistasis parameter (see
 (\ref{Qss})) and as a consequence, the fixation time is also
 insensitive to epistasis. But in regime II, the mutator fraction exhibits
 a nontrivial dependence on epistasis (see (\ref{Qws})) and decreases
 as epistasis changes from antagonistic to synergistic. As a result,
 the fixation time for  
 synergistically  interacting fitnesses is larger than that for
 antagonistic ones. The implications of our results for the evolution
 of mutation  rates are discussed.

%============================================================================
%MODEL
%============================================================================

\section*{Models}

We consider a haploid, asexual population evolving on a 
fitness landscape in which the Malthusian fitness $F(k)$ of
a genotype carrying $k$ deleterious mutations 
is given by \citep{Wiehe:1997}
\be
F(k) = -s {k^\alpha}
\label{fitness}
\ee
where $s$ is the cost of deleterious mutation and the parameter
$\alpha \geq 0$ controls epistasis. If $\alpha=1$, there is no
epistasis while 
$\alpha < 1$ and $> 1$ respectively correspond to antagonistic
and synergistic epistasis.  A deleterious mutation occurs at a rate
$U$ for a nonmutator and  $V=\lambda
U, \lambda \gg 1$ for a mutator.  Besides replication and mutation, a 
mutator is created  
from a nonmutator at a forward conversion rate $f$ but the reverse reaction  
is ignored since such events occur at a backward rate $b \ll f$ 
\citep{Ninio:1991,Boe:2000}. 

We implement the stochastic dynamics of a population of fixed
  size $N$ using the standard Wright-Fisher process in which  
each individual in the new generation chooses a parent with
a probability proportional to the parent's fitness. Then the  mutations 
chosen from a Poisson distribution  
with mean given by the  
parent's mutation rate are introduced and finally a 
nonmutator offspring is converted into a mutator.

In an infinite population in which the time evolution occurs 
deterministically, the average fraction ${\cal P}(k,t)$ and
${\cal Q}(k,t)$ of 
  nonmutators and mutators 
  respectively with $k$ deleterious mutations at time $t$ 
  evolves according to the following coupled equations:
\bea
\frac{\partial {\cal P}(k,t)}{\partial t} &=& U ({\cal P}(k-1,t)-{\cal
  P}(k,t))+ (F(k)-{\cal W}(t)) {\cal P}(k,t)- f {\cal P}(k,t) \label{Pkt}
\\
\frac{\partial {\cal Q}(k,t)}{\partial t} &=& V ({\cal Q}(k-1,t)-{\cal
  Q}(k,t))+ (F(k)-{\cal W}(t)) {\cal Q}(k,t)+f {\cal P}(k,t)  \label{Qkt}
\eea
In the above equations, the fractions ${\cal P}(-1,t)={\cal Q}(-1,t)=0$ and 
\be
{\cal W}(t) = \sum_{k=0}^{\infty} F(k) ({\cal P}(k,t)+{\cal Q}(k,t))
\ee
is the population fitness 
which ensures that the normalisation condition $\sum_{k=0}^\infty {\cal
  P}(k,t)+ {\cal Q}(k,t)=1$ is satisfied in all generations. 
The above equations can be derived using standard methods
\citep{Baake:2000} starting 
from a discrete time model \citep{Nagar:2009}.

%============================================================================
%RESULTS
%============================================================================

\section*{Results}

%___________________________________________________________________________

\subsection*{FIXATION OF MUTATORS IN INFINITE AND FINITE POPULATIONS}

\begin{figure}
\begin{center}
\includegraphics[width=0.75 \linewidth,angle=270]{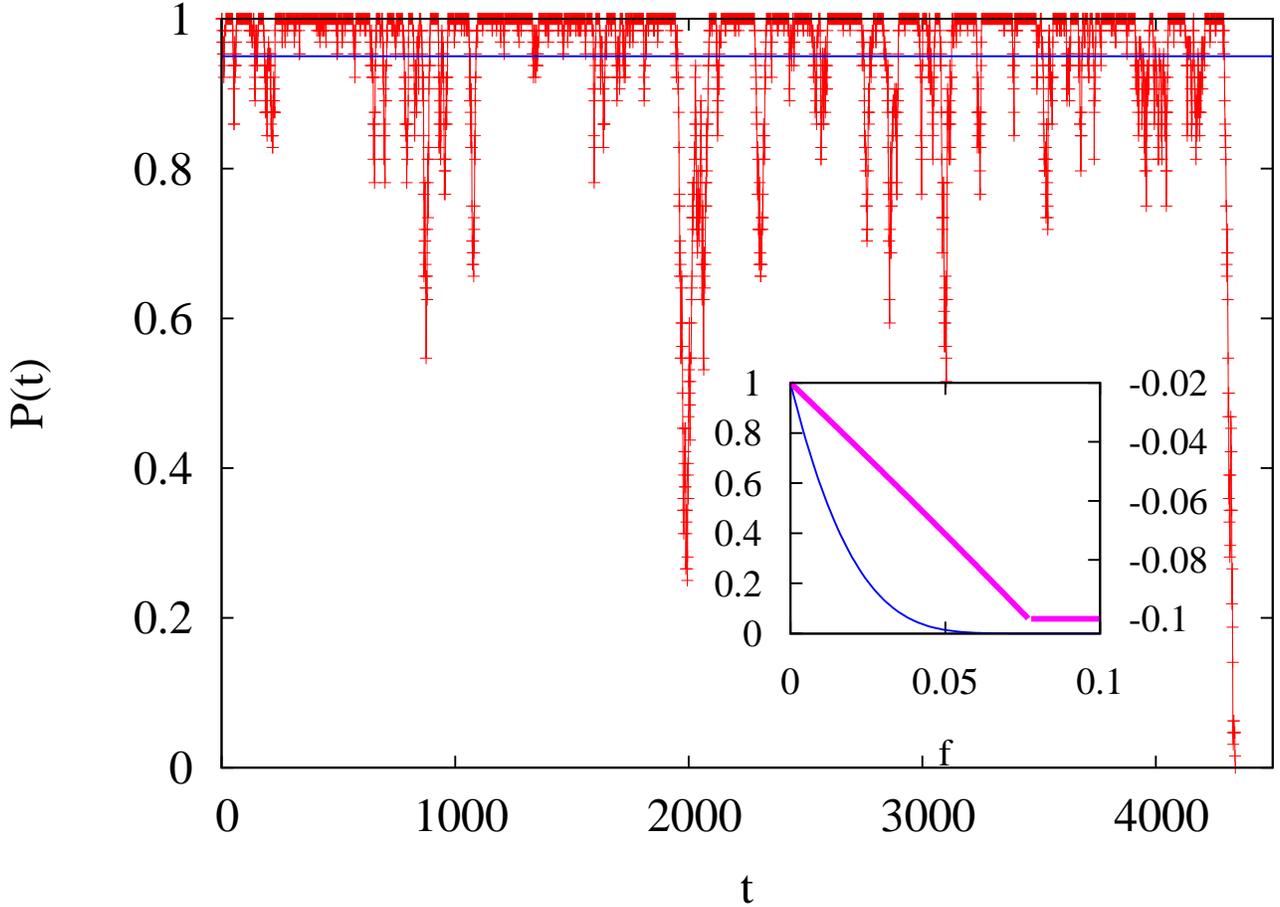}
\caption{Loss of nonmutators in a finite population (main) 
     due to 
  stochastic fluctuations and continual production of mutators at a 
  rate $f=10^{-3}$ and in an infinitely large 
  population (inset) at a critical conversion rate $f_c$. The 
   horizontal line in the main figure is the nonmutator
   fraction at mutation-selection 
  balance. This fraction is shown in the inset on the left
    y-axis as a function of $f$. The inset also shows the mean fitness
    ${\cal W}$ 
    (thick line) 
    for an infinitely large population on the right y-axis. The
    other parameters used in all plots are  $s=10^{-2},U=2 \times
    10^{-2} ,\lambda=5, \alpha=1$.}
\label{run_fig}
\end{center}
\end{figure}

We first show that in an infinite population
at mutation-selection 
balance, a minimum conversion rate $f_c$ is required to fix the
mutators \citep{Tannenbaum:2005,Nagar:2009}. 
Although the nonmutator population carries less mutational load than
the mutators, it suffers a loss due to 
the creation of mutators when the rate $f$ is nonzero. 
As a result, there is a transition at a critical conversion rate $f_c$ 
below which the population is in a {\it mixed state} composed
of both nonmutator and mutator subpopulations and above which only
mutators are present (see inset of Fig.~\ref{run_fig}). 

To see this, consider the nonmutator and mutator frequencies at 
mutation-selection 
balance when they become time-independent. Then (\ref{Pkt}) and
(\ref{Qkt}) reduce to 
\bea
U {\cal P}(k-1)+ (F(k)-{\cal W}-f-U) {\cal P}(k) &=& 0 \label{Pkss}\\
V {\cal Q}(k-1)+ (F(k)-{\cal W}-V) {\cal Q}(k)+f {\cal P}(k) &=& 0 \label{Qkss}
\eea
\noindent{\it Mixed state:} It is clear from (\ref{Pkss}) that
a nonzero distribution of nonmutators  is possible only
if the fraction ${\cal P}(0)$  is
nonzero.  For $k=0$ in (\ref{Pkss}), we immediately have 
\be
{\cal W}=-f-U ~,~{\cal P}(0) \neq 0
\label{ssW}
\ee
Thus the mean fitness in the mixed state is independent of
the epistasis 
  exponent $\alpha$, mutator rate $V$ and selection effect 
  $s$. This result is a simple generalisation of the well known result 
  for $f=0$ \citep{Kimura:1966,Haigh:1978}. 

\noindent{\it Pure mutator state:} If the nonmutator
  frequency in the fittest class is zero, (\ref{Qkss}) reduces to
\be
V {\cal {\tilde Q}}(k-1)+ (F(k)-{\cal {\widetilde W}}-V) {\cal {\tilde
    Q}}(k) = 0
\label{one}
\ee
where a tilde is used to refer to quantities in the pure
  mutator state. 
Equation (\ref{one}) and its relatives have been studied extensively
\citep{Jain:2007b,Jain:2011c} and it has been shown that  
an error threshold transition \citep{Eigen:1971} can occur in which a 
population localised around the fittest sequence gets delocalised
beyond a critical mutation rate. For the fitness function
(\ref{fitness}), such a transition occurs in the special case of 
sharp peak fitness landscape ($\alpha \to 0$) but the 
population remains in the localised state at any mutation rate for
$\alpha > 0$ \citep{Wiehe:1997,Baake:2000}. Since the population
fraction in the fittest class is nonzero in the localised state, using
(\ref{one}) for $k=0$, we get 
\be
{\cal {\widetilde W}}=-V ~,~{\cal {\tilde Q}}(0) \neq 0
\label{ssWone}
\ee

\noindent{\it Critical conversion rate:} As shown in the
  inset of Fig.~\ref{run_fig}, with increasing rate $f$, 
the mixed state fitness ${\cal W}$ decreases linearly until
it reaches the pure mutator state fitness ${\cal {\widetilde W}}$
\citep{Nagar:2009}. 
Matching the mean fitnesses (\ref{ssW}) and (\ref{ssWone})
in the two states gives the critical forward conversion rate as  
\be
f_c=V-U
\label{fcdefn}
\ee
which does not depend on the epistasis parameter $\alpha$. (However for sharp
peak fitness landscape, an additional critical 
conversion rate exists besides (\ref{fcdefn}), which is discussed in
Appendix A.)

While a minimum conversion rate is
required to fix mutators in an infinitely large population, the
mutators get fixed for any $f > 0$ when the population is of finite
size. 
Starting with a
population of $N$ nonmutators each with fitness one, we follow 
the dynamics of the total nonmutator frequency 
$P(t)=\sum_{k=0}^\infty P(k,t)$ where $P(k,t)$ is the instantaneous
fraction in the fitness class $k$. 
As shown in Fig.~\ref{run_fig}, the fraction $P(t)$ first relaxes to a
  value close to the 
  deterministic nonmutator fraction 
  ${\cal P}=\sum_{k=0}^\infty {\cal 
  P}(k)$ (phase 1) about which it fluctuates (phase 2) until it enters the
  absorbing state with zero nonmutators (final state).

%============================================================================

\subsection*{TIME TO FIX MUTATORS IN A FINITE POPULATION}

Ignoring the time for the nonmutator population to relax 
  to its deterministic frequency,  
  we  compute the time for the loss of nonmutators 
  starting from deterministic fraction ${\cal
  P}$ using a diffusion theory \citep{Ewens:1979}. We begin by
  calculating the mutator fraction at mutation-selection balance as it
  is required to find the fixation time. In the following,
  we work in the biologically relevant limit 
  $f \to 0$ \citep{Ninio:1991,Boe:2000} and also assume that the
  epistasis exponent is nonzero and finite ($0 < \alpha < \infty$).

%============================================================================

\subsubsection*{Mutator fraction at mutation-selection balance}

To calculate the mutator fraction, we first expand the population
fractions in a power series about $f=0$ by writing 
\bea
{\cal P}(k;f) &=& \sum_{n=0}^\infty f^n \frac{{\cal P}_n(k)}{n !} \label{pertP} \\
{\cal Q}(k;f) &=& \sum_{n=0}^\infty f^n \frac{{\cal Q}_n(k)}{n!}
\label{pertQ}
\eea
where $P_n(k)$ and $Q_n(k)$ respectively denote the $n$th derivative
of the nonmutator and mutator fraction with respect to rate $f$
evaluated at $f=0$. Since ${\cal Q}_0(k)={\cal Q}(k;f=0)=0$
for all $k$ \citep{Nagar:2009}, the mutator fraction ${\cal Q}(k;f)
\approx f {\cal Q}_1(k)$ where we have neglected terms of order
$f^2$ and higher in (\ref{pertQ}).  On using (\ref{ssW}),
(\ref{pertP}) and (\ref{pertQ}) in the steady state equation (\ref{Qkss}), 
a straightforward 
calculation shows that ${\cal Q}_1(k)$ obeys the following equation:
\bea
V {\cal Q}_1(k-1)+ (F(k)-\delta U) {\cal Q}_1(k)+ {\cal P}_0(k) = 0
\label{Qkssp} 
\eea
where $\delta U=V-U$. Similarly using (\ref{Pkss}),(\ref{ssW}) and
(\ref{pertP}),  
the equation for nonmutator fraction ${\cal P}_0(k)$ can be
written as 
\be
U {\cal P}_0(k-1)+ F(k) {\cal P}_0(k) = 0 \label{Pkssp}
\ee
which can be readily iterated to yield 
\bea
{\cal P}_0(k) =\left(\frac{U}{s}\right)^k \frac{ {\cal
    P}_0(0)}{(k!)^\alpha} 
\label{Pksoln}
\eea
where the fraction ${\cal P}_0(0)$ is given by
\be
{\cal P}_0(0)=\left[\sum_{j=0}^\infty \left(\frac{U}{s}\right)^j
  \frac{1}{(j!)^\alpha} \right]^{-1} 
\ee
The sum on the right hand side (RHS) of the above equation does not
appear to be known in terms of standard mathematical functions except
for some special values of $\alpha$ such as $1$ and $2$
\citep{Jain:2008b}. For the mutator fractions, (\ref{Qkssp}) gives 
\be
{\cal Q}_1(k)= \frac{V~{\cal Q}_1(k-1)}{\delta U-F(k)} + \frac{{\cal
    P}_0(k) }{\delta U-F(k)} 
\label{Qkssp2}
\ee
 On using ${\cal Q}_1(-1)=0$ in the above equation
for $k=0$, we get 
\be
{\cal Q}_1(0)=\frac{{\cal P}_0(0)}{\delta U} 
\label{Q10}
\ee
For other fitness classes, on repeatedly
iterating (\ref{Qkssp2}), we find that 
\be
{\cal Q}_1(k)= {\cal Q}_1(0) \prod_{m=1}^k \frac{V}{\delta U-F(m)} +\frac{1}{V}
\sum_{j=1}^k {\cal P}_0(j) \prod_{m=j}^k \frac{V}{\delta U-F(m)}~,~k > 0
\ee
Finally on summing over all the fitness classes, we obtain the
mutator fraction to leading order in $f$ as
\bea
{\cal Q} &=& \frac{f}{V} \sum_{k=0}^\infty \sum_{j=0}^k {\cal P}_0(j)
\prod_{m=j}^k \frac{V}{\delta U+s m^\alpha} \\
&=&  \sum_{k=0}^\infty \sum_{j=0}^k {\cal P}_0(j)
\frac{G(k)}{G(j)} \frac{f}{\delta U+s j^\alpha} 
\label{Qcont}
\eea
where we have defined 
\be
G(k) = \prod_{m=1}^k \frac{V}{\delta U+s m^\alpha} 
\label{prod}
\ee

To evaluate the mutator fraction in (\ref{Qcont}), we first estimate
the product $G(k)$ as explained in Appendix B. Using the result
(\ref{prodestim}) in (\ref{Qcont}), we find that  
\bea
{\cal Q} \approx \frac{f}{\delta U} \sum_{k=0}^\infty \sum_{j=0}^k
\left( \frac{V}{\delta U} \right)^{k-j} {\rm exp}\left[{-\frac{s}{\delta U}
  \frac{(k^{\alpha+1}-j^{\alpha+1})}{\alpha+1}} \right]{\cal P}_0(j)
\label{Qcont2}
\eea

%--------------------------------------------------------------------------

\noindent{\it Regime I:} If $s/\delta U \gg 1$, the dominant
contribution to the  inner sum on the RHS of
(\ref{Qcont2}) comes from the term with $j=k$. Neglecting the summand
with $j \neq k$, we immediately obtain  
\be
{\cal Q} \approx \frac{f}{\delta U}~,~\delta U/s \ll 1
\label{Qss}
\ee
Note that this solution is independent of the selective cost $s$ and
the epistasis exponent $\alpha$.

%--------------------------------------------------------------------------

\noindent{\it Regime II:} For $s/\delta U \ll 1$, the evaluation of
the double sum in (\ref{Qcont2}) is nontrivial and 
we calculate the mutator fraction in this regime within a Gaussian
approximation as explained in Appendix C. 
Plugging (\ref{P0k}) and (\ref{Gk}) in the expression (\ref{Qcont})
for mutator fraction and approximating the sums by integrals, we get  
\bea
{\cal Q} &\approx&  \frac{f}{\delta U} \sqrt{\frac{\alpha}{2 \pi k^*}}
\int_{0}^\infty dk ~e^{- \frac{\alpha U (k-k^*)^2}{2 k^* \delta 
    U}} \int_{0}^k dj~e^{ \frac{\alpha U (j-k^*)^2}{2 k^* \delta
    U}} ~e^{-\frac{\alpha
    (j-k^*)^2}{2 k^*}} \\
&\approx & \frac{f}{2 \delta U}  \int_{0}^\infty dk ~e^{-
  \frac{\alpha U (k-k^*)^2}{2 k^* \delta 
    U}} \left[ {\rm erf}(\sqrt{k^*}) + \mathrm{erf} \left(\sqrt{\frac{\alpha}{2 k^*}} (k-k^*) 
  \right) \right] 
\eea
where ${\rm erf}(x)$ is the error function. 
For $k^*, \lambda \gg 1$, the second integral on the RHS of the above
equation vanishes and the first integral can be readily evaluated 
to give 
\bea
{\cal Q} &\approx& f \sqrt{\frac{\pi}{2 \alpha} \frac{U^{\frac{1-2
	\alpha}{\alpha}}}{(\lambda-1) s^{\frac{1}{\alpha}}}}~,~\delta U/s \gg 1
\label{Qws}
\eea
For $\alpha=1$, we recover the known results \citep{Nagar:2009,Desai:2011}.  

%--------------------------------------------------------------------------

Our analytical results (\ref{Qss}) and (\ref{Qws}) are compared with 
the exact expression (\ref{Qcont}) evaluated numerically in 
Figs.~\ref{Q_s}-\ref{Q_lambda} and show a good agreement. A detailed
discussion of these results is given in the last section of this
article.

%============================================================================

\subsubsection*{Diffusion approximation for fixation time}

We now develop a 
diffusion theory \citep{Ewens:1979}
to estimate the average  time to lose all the
nonmutators.  Using a backward Fokker-Planck equation, it can be shown
that the 
average absorption time ${\overline T}$ to the state with zero
nonmutators starting from total nonmutator fraction 
${{\cal P}}$  is given by \citep{Ewens:1979,Jain:2008b} 
\bea
{\overline T}
&=& 2 \int_0^{{\cal P}} dy ~\psi(y) \int_y^1 \frac{dx}{D_2(x) \psi(x)} 
\label{Tdefn}
\eea
where the function 
\be
\psi(x)= e^{-2 \int dx \frac{D_1(x)}{D_2(x)}}
\label{psi}
\ee
and $D_1(x), D_2(x)$ are  drift and diffusion coefficients respectively.

To calculate an expression for the drift coefficient $D_1$, we first
sum over all $k \geq 0$ in (\ref{Pkt}) and find that 
\be
\frac{\partial {\cal P}(t)}{\partial t}={\cal W}_P(t)- (f+{\cal W}(t))
     {\cal P}(t)
\label{WPt}
\ee
where   ${\cal P}(t)=\sum_{k=0}^\infty {\cal P}(k,t)$  and ${\cal
  W}_P(t)=\sum_{k=0}^\infty F(k) {\cal P}(k,t)$.  At
mutation-selection balance, due to (\ref{ssW}) 
 and (\ref{WPt}), the  mean fitness of the nonmutator
 population is given by   
\be
{\cal W}_P=-U {\cal P}
\label{WPgenl}
\ee
Using the definition of drift coefficient
\citep{Ewens:1979,Jain:2008b} and the result (\ref{WPt}), we may write
\bea
D_1(P(t) = {\cal P}(t)) &=& \frac{\partial {\cal P}(t)}{\partial
  t} \\
&=& P \times \left(\frac{W_P}{P}- f-W \right) 
\label{open}
\eea
Since the above equation does not close in $P(t)$, 
we use a simple 
argument to find an approximate expression for $D_1$
\citep{Stephan:1993,Stephan:2002}. We first note that $D_1$ has two
zeros namely $0$ and ${\cal P}$ since the nonmutator fraction $P(t)$ does
not change with time in the absorbing state and 
(quasi)steady state. Therefore we can write 
\be
D_1 \approx C P ~\left(1- \frac{P}{{\cal P}}\right)
\label{close}
\ee
where $C$ is a proportionality constant which can be
determined by considering the second factor on the RHS
of (\ref{open}) when the average population fraction is close to the 
mutation-selection balance (phase 2). Assuming that the
population fitnesses $W_P(t)$ and $W(t)$ 
have the same functional form in phase 2 as in the time-independent
steady state, we can express them in terms of $P(t)$ as
\bea
W_P(P) & \approx & -U P \\
W(P) & \approx & -U -\frac{1-P}{{\cal Q}_1}
\eea
where we have used (\ref{WPgenl}) and eliminated $f$ from (\ref{ssW})
in favor of ${\cal P}$ 
using ${\cal Q}=f {\cal Q}_1$. A power series expansion of the
preceding equations in $P-{\cal P}$ gives
\bea
\frac{W_P}{P}- f-W &\approx& 
\left(\frac{1}{{\cal Q}_1}-f \right)~\left(1- \frac{P}{{\cal P}}\right)
\eea
On using the last expression in (\ref{open}) and comparing it with
(\ref{close}), the constant $C$ is immediately determined.

Furthermore as the number of offspring produced per generation are distributed
according to a binomial distribution, the variance in the offspring
number gives \citep{Jain:2008b}
\be
D_2= \frac{P (1-P)}{N}
\label{D2}
\ee

\begin{figure}
\begin{center}
\includegraphics[width=0.6 \linewidth,angle=270]{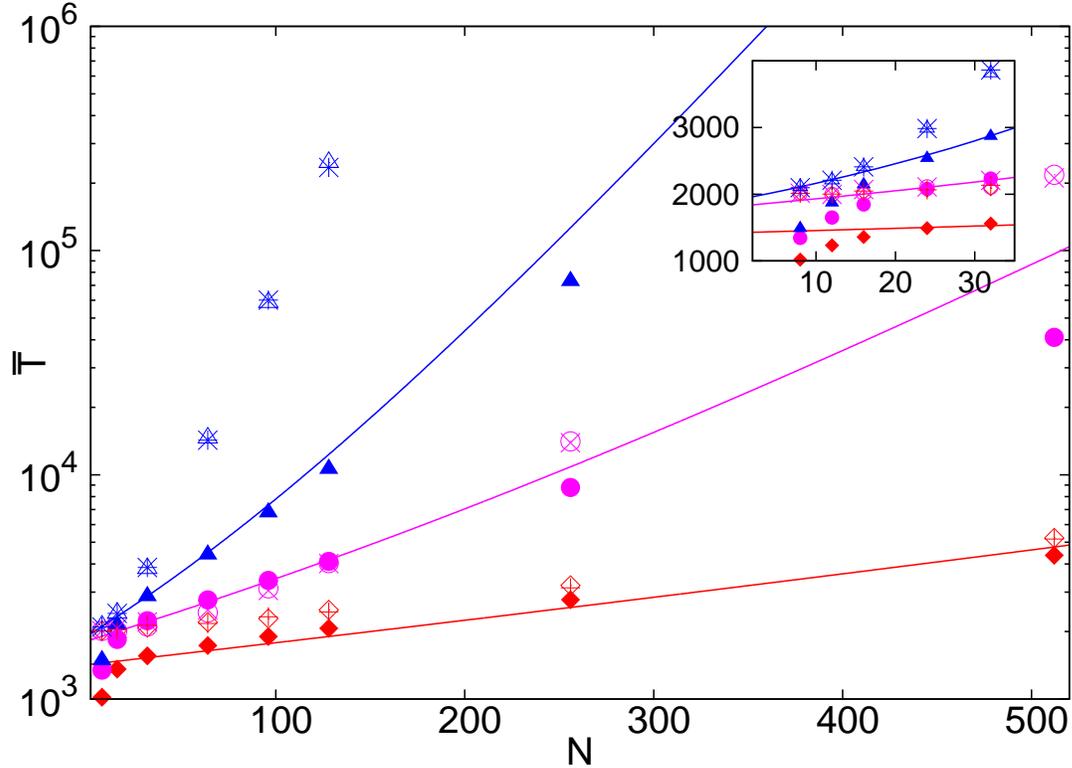}
\caption{Variation of fixation time with population size $N$ when
  mutational effects are stronger than the selection (regime II).  The
  simulation data is 
  obtained  by averaging over $10^4$ independent
  realisations for $\alpha=1/2(\diamond, +), 1(\circ, \times), 2
  (\vartriangle,\ast)$ for $b=0$
  and $0.01 f$ respectively. The filled symbols 
  show the result 
  of numerical integration of (\ref{Tapp}) while the solid
  curves are the analytic formulae (\ref{TJfix}). The other parameters are 
$f=0.0005,~s=0.001,~U=0.005$ and $\lambda=10$.  The inset shows the
  data for the same parameters as in the main figure for small
  populations. The simulation data for $\alpha=2$ does not match
  quantitatively with the diffusion theory possibly because
  $k^*=(U/s)^{1/\alpha} \approx 6.7$ is rather small.}
\label{TN2}
\end{center}
\end{figure}

Using the above expressions for $D_1$ and
$D_2$ in (\ref{Tdefn}), the 
fixation time for mutators can  
be estimated as described in Appendix D. The final result is 
\be
{\overline T} =
\begin{cases}
\frac{{\cal Q}}{2 N f^2} \left(e^{\frac{2 N f}{{\cal
      Q}}}- e^{2 N f} \right)~,~ N \ll 1/(2 f) \\
\frac{{\cal Q}^2}{2 N f^2 {\cal P}^2} e^{\frac{2 N f {\cal P}}{{\cal
      Q}}}~~~,~ N \gg 1/(2 f)
\end{cases}
\label{TJfix}
\ee
where the mutator fraction ${\cal Q}$ is given by (\ref{Qss}) or
(\ref{Qws}). 
The fixation time obtained using the above expressions and numerical
simulations is shown in 
Figs.~\ref{TN1} and \ref{TN2} for representative values of
$\alpha$. The following section discusses these results in more
detail.

%============================================================================
%DISCUSSION
%==========================================================================

\section*{Discussion}

In this article, we studied the dynamics of a population of
nonmutators and mutators, starting from a well adapted population of
nonmutators, when the mutations are unconditionally deleterious.

\subsection*{Effect of genetic drift} 

The nonmutator population decreases due to a conversion
into mutator genotypes. But in an infinitely large
population, this loss can be compensated if the conversion rate  
is sufficiently small. On a class of fitness landscapes defined by
(\ref{fitness}), a transition occurs at 
a critical rate $f_c$ between a mixed state with 
nonmutators and mutators and a pure mutator state
\citep{Tannenbaum:2005,Nagar:2009}. Interestingly, the 
critical conversion rate given by (\ref{fcdefn}) is
independent of $\alpha$ (but see Appendix A also). As the previous
studies \citep{Johnson:1999b,Desai:2011} on infinite populations 
worked exclusively in the small $f$ limit, 
the possibility of a transition was
not realised. Using $U=2 \times 10^{-3}$ \citep{Drake:1998,Boe:2000}
and $\lambda=10$ in (\ref{fcdefn}), we find that the transition
occcurs at $f_c \sim 10^{-2}$ which is several orders of magnitude 
larger than the spontaneous conversion rate  $f\sim 10^{-6}-10^{-7}$ 
estimated in {\it E. coli} \citep{Ninio:1991,Boe:2000}.

The difference between an infinite and finite population is that 
in the latter case, the nonmutators get wiped out for arbitrarily
small $f$ in a finite time as illustrated in
  Fig.~\ref{run_fig}. If the conversion rate is set  
to zero in a finite population of nonmutators
and mutators,  either of them can fix
\citep{Wylie:2009}. But for nonzero $f$, due to the  simple fact that 
the mutators are being continually generated, a finite population of 
nonmutators will be converted into mutators in a finite time.

\subsection*{Evolutionary regimes}

\begin{figure}
\begin{center}
\includegraphics[width=0.75 \linewidth,angle=270]{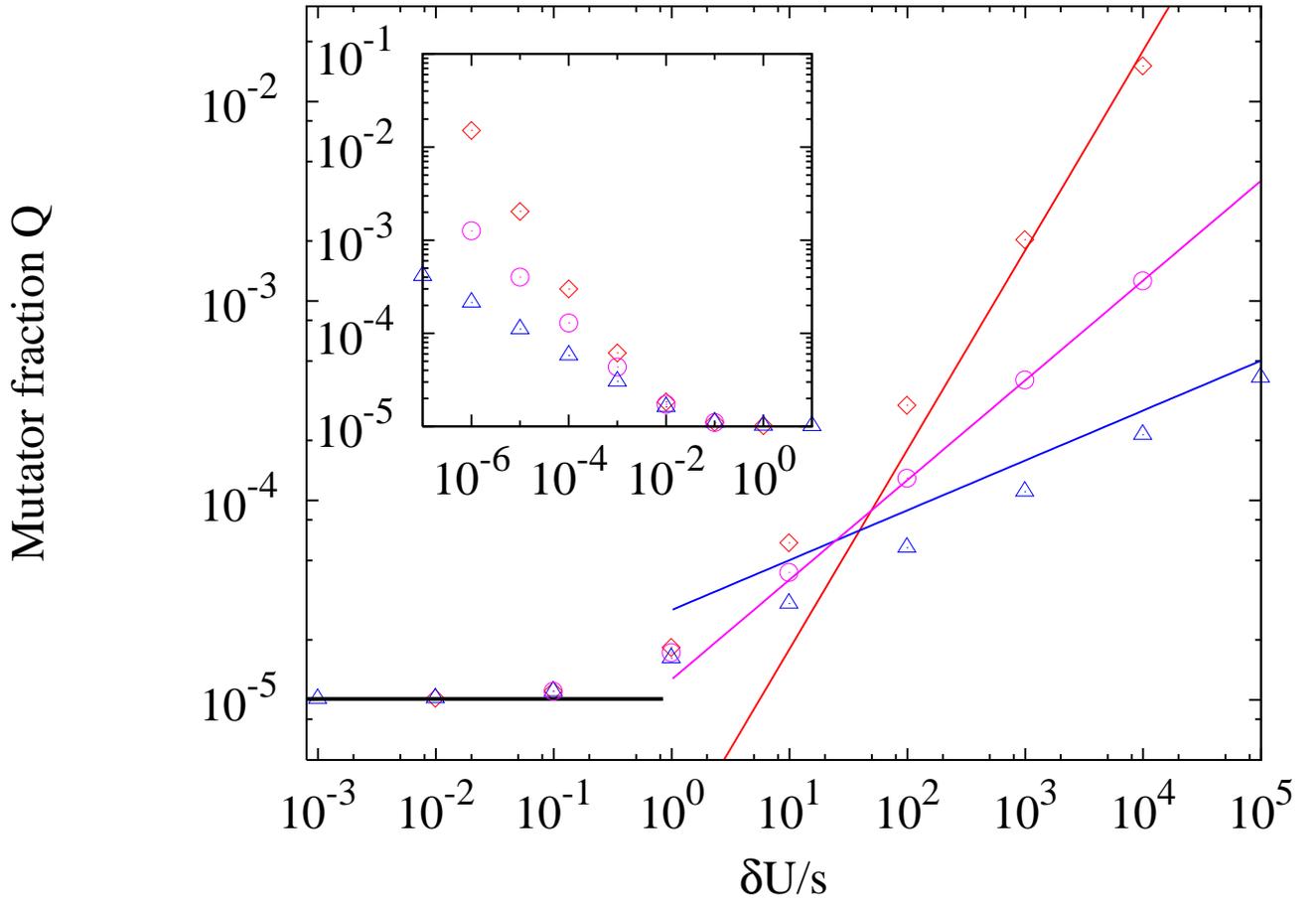}
\caption{Variation of mutator fraction at mutation-selection balance
  with selective cost $s$ for $\alpha=1/2 
  (\diamond), 1(\circ), 2 (\vartriangle)$. The other parameters are
  $f=10^{-7},   
  U=10^{-4},\lambda=10^2$. The exact sum (\ref{Qcont}) is shown by
  points and the asymptotic expressions (\ref{Qss}) and (\ref{Qws}) by
  solid curves. The inset shows the data in the main figure when
  plotted as a function of selective effect $s$.} 
\label{Q_s}
\end{center}
\end{figure}

\begin{figure}
\begin{center}
\includegraphics[width=0.75 \linewidth,angle=270]{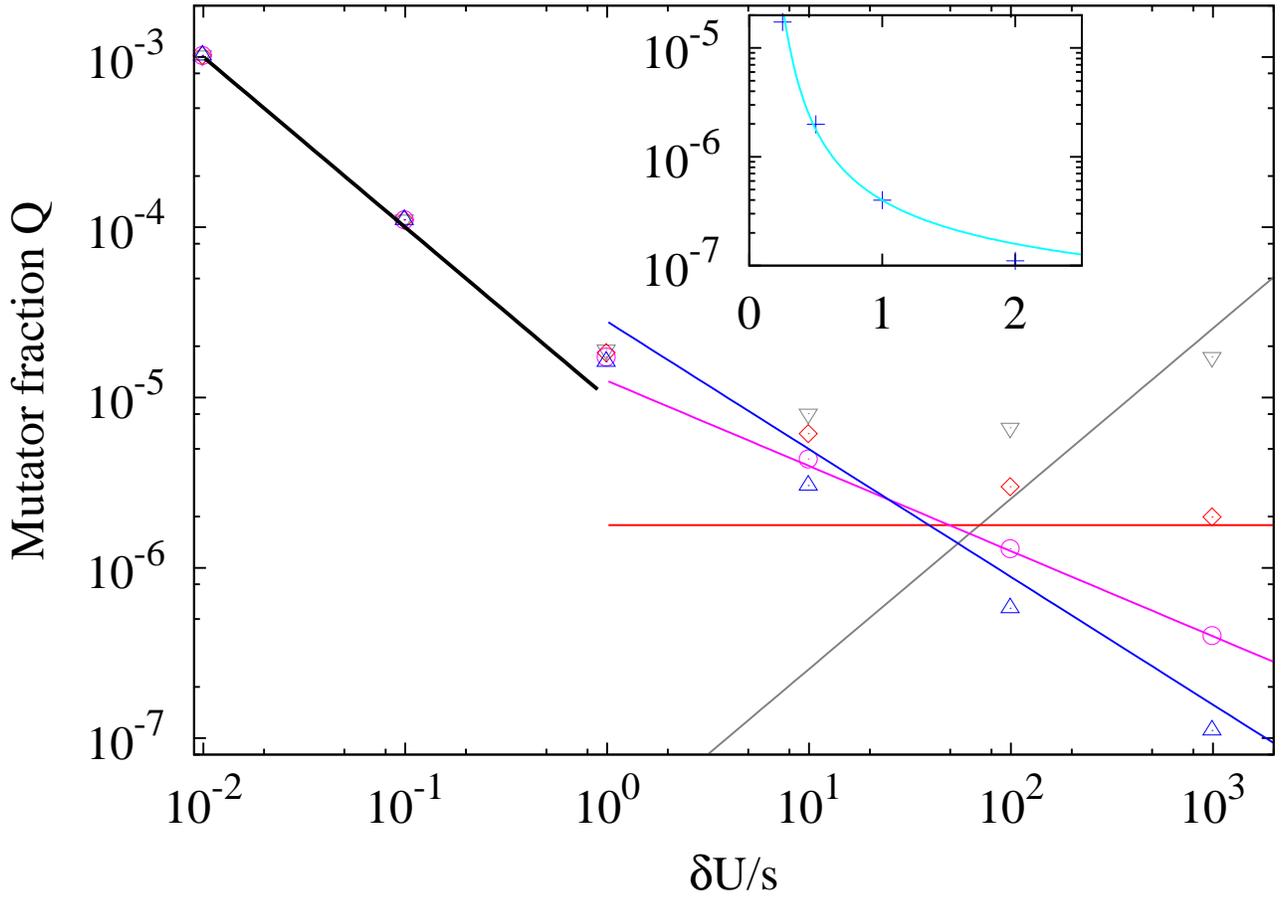}
\caption{Variation of mutator fraction at mutation-selection balance
  with mutation rate $U$ for $\alpha=1/4(\triangledown), 1/2
  (\diamond), 1(\circ), 2 (\vartriangle)$. The other parameters 
are $f=10^{-7},  
  s=10^{-2},\lambda=10^2$. The exact sum (\ref{Qcont}) is shown by
  points and the asymptotic expressions (\ref{Qss}) and (\ref{Qws}) by
  solid curves. The inset shows the data in the main figure
    when plotted against the exponent $\alpha$ at fixed $\delta U/s=990$.}
\label{Q_U}
\end{center}
\end{figure}

\begin{figure}
\begin{center}
\includegraphics[width=0.75 \linewidth,angle=270]{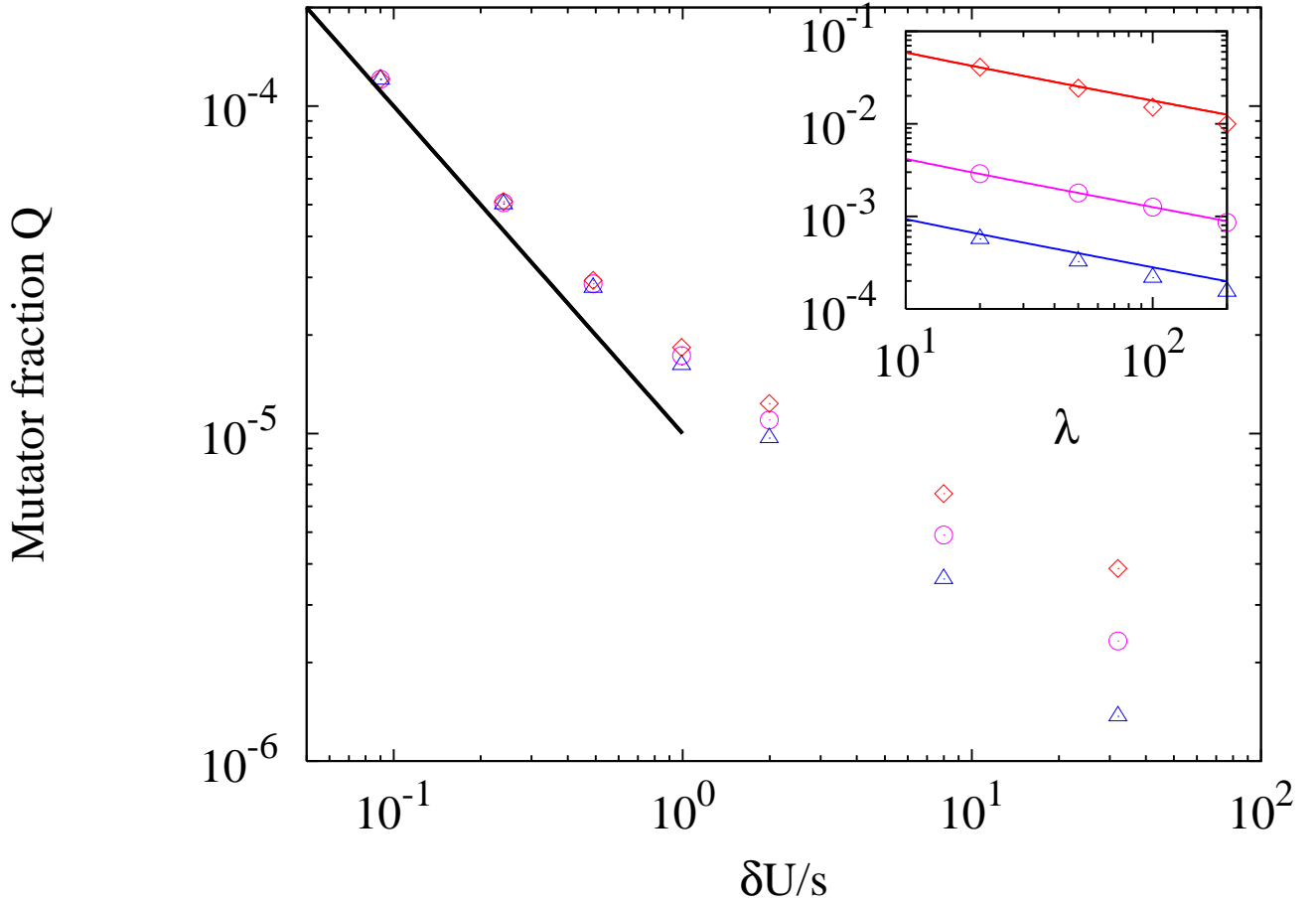}
\caption{Variation of mutator fraction at mutation-selection balance
  with mutator strength $\lambda$ for $\alpha=1/2
  (\diamond), 1(\circ), 2 (\vartriangle)$. The selective cost $s$ is 
  $10^{-2}$ (main), $10^{-6}$ (inset) while the other parameters are
  $f=10^{-7}, U=10^{-4}$.  The exact sum
  (\ref{Qcont}) is shown by  
  points and the asymptotic expressions (\ref{Qss}) and (\ref{Qws}) by
  solid curves. The result (\ref{Qws}) is not shown in the main figure 
  as it holds when $k^*=(U/s)^{1/\alpha}\gg 1$ but it is compared with
  the exact expression (\ref{Qcont}) in the inset.}
\label{Q_lambda}
\end{center}
\end{figure}

The mutator fraction at
mutation-selection balance has been calculated by several
authors on 
non-epistatic fitness landscapes. While
\citet{Johnson:1999b,Desai:2011} computed it for a general
distribution of fitness effects and small conversion rate 
$f$, an exact solution was obtained for arbitrary $f$ by
\citet{Nagar:2009} when all mutations have the same selective
effect. In their work, 
\citet{Desai:2011} have pointed  out that the behavior of the mutator
fraction depends on the relative strength of mutational effects
($\delta U$) to selective effects ($s$).  
When selective disadvantage exceeds the mutational effects, the
mutations can be treated as effectively lethal \citep{Johnson:1999b,
  Desai:2011}. This is because when $\delta U
\ll s$ (regime I), the most populated nonmutator fitness class $k^*$
in (\ref{kstar}) is below unity and therefore the  
nonmutator population is effectively localised in the fitness class with
zero deleterious mutations.  As a consequence, mutators which 
arise due to nonmutators, are also localised in the zero mutation
class. Setting all the nonmutator and mutator fractions  
other than in the zeroth mutational class to zero, (\ref{Q10})  
immediately leads to (\ref{Qss}) which is (unsurprisingly) independent
of epistasis. Moreover the result (\ref{Qss}) is 
  expected to be 
  insensitive to the distribution of selective effects also 
  \citep{Desai:2011} since both subpopulations are
  localised in the zero mutation class. However in the 
opposite limit $\delta U \gg s$ (regime II), mutator
  subpopulation is spread over 
many fitness classes and the simple argument given above
underestimates the mutator fraction  as can be seen by
extrapolating the solution (\ref{Qss}) to regime II in
Figs.~\ref{Q_s}-\ref{Q_lambda}. An expression for mutator frequency in 
regime II has been 
obtained in the absence of epistasis
\citep{Johnson:1999b,Nagar:2009,Desai:2011} and here we have extended
the previous analyses by including it.  

Figures
\ref{Q_s}-\ref{Q_lambda} show that the mutator fraction in 
regime II  decreases monotonically as the epistasis parameter $\alpha$  
increases. To understand this behavior, 
we first note that  the fitness difference $F(k)-F(k+1) \approx s
\alpha k^{\alpha-1}$ for large $k$ approaches zero for $\alpha <
1$ and infinity for $\alpha > 1$ as $k$ increases. 
Thus the mutators stand a better chance of 
survival when the fitness interactions are antagonistic than
synergistic. For 
a given $\alpha$, as large selective cost is detrimental for the mutators, the
mutator frequency decreases (regime II) to a constant (regime I) as $s$ 
increases (refer Fig.~\ref{Q_s}). 
The effect of increasing mutator strength $\lambda$ is also 
to reduce the mutator fraction as shown in Fig.~\ref{Q_lambda}. 
 The variation of mutator frequency with the spontaneous mutation rate
$U$ shown in Fig.~\ref{Q_U} however displays an interesting feature:
 as $U$ increases, the mutator frequency decreases monotonically for
 $\alpha > 1/2$ but for $\alpha < 1/2$, it decreases in
regime I and increases in regime II. At
$\alpha=1/2$, the mutator fraction decreases to a constant as mutation 
rate $U$ is raised. This behavior may be understood as follows: 
 in regime II in which both subpopulations are spread over
the fitness landscape and carry many deleterious mutations, both nonmutators
and mutators incur a negligible fitness cost when a mutation occurs
for small $\alpha$ but due to the unidirectional flux from 
nonmutators to mutators, the mutator fraction approaches unity as
$U$ increases (also see Appendix A). This suggests that even if the spontaneous 
mutation rate of the nonmutator is  high, the mutators can still thrive
provided the fitness cost increases sufficiently slowly.

\begin{figure}
\begin{center}
\includegraphics[width=0.6 \linewidth,angle=270]{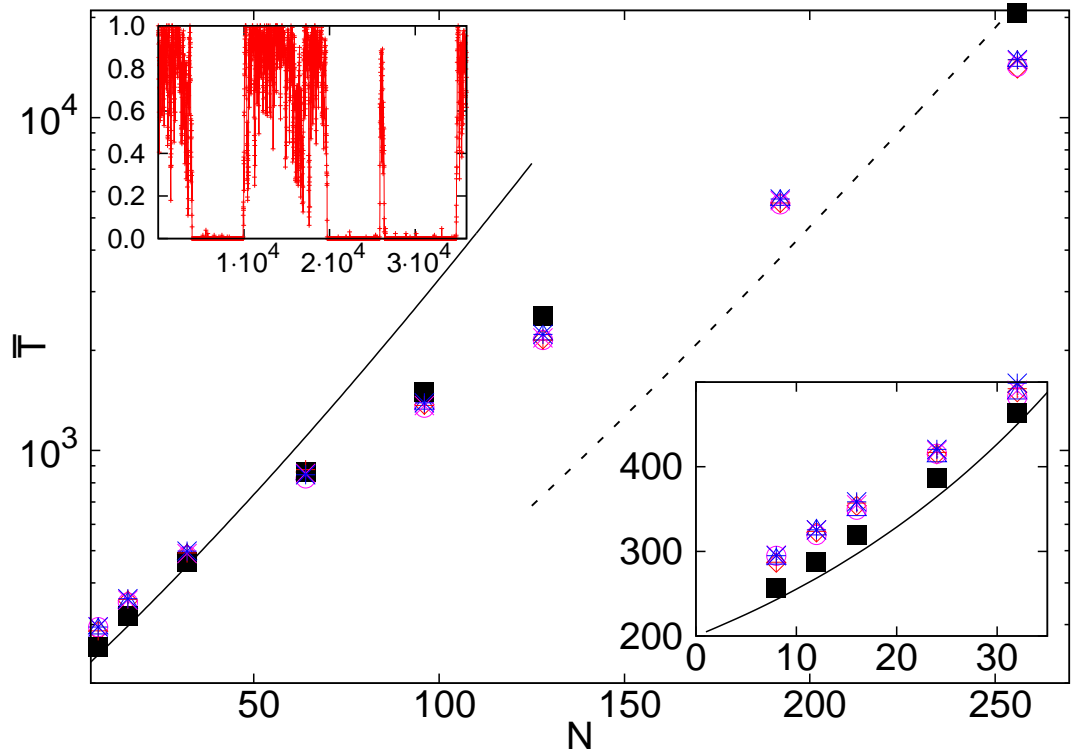}
\caption{Variation of fixation time with population size $N$ when
  mutational effects are weaker than the selection (regime I). The
  simulation data is 
  obtained  by averaging over $10^4$ independent
  realisations for $\alpha=1/2(\diamond, +), 1(\circ, \times), 2
  (\vartriangle,\ast)$ for $b=0$
  and $0.01 f$ respectively. The points 
  shown by filled squares are the result 
  of numerical 
  integration of (\ref{Tapp})  while the solid and dashed 
  curves are the analytic formulae (\ref{TJfix}).  The other
  parameters are $f=0.004,~s=0.2,~U=0.005$ and 
  $\lambda=5$.  The right bottom inset shows the 
  data for the same parameters as in the main figure for small
  populations and the left top inset shows the nonmutator fraction as
  a function of time for $N=128$ and $\alpha=1$.}
\label{TN1}
\end{center} 
\end{figure}

On non-epistatic fitness landscapes, the short
time dynamics \citep{Johnson:1999b,Desai:2011} until the 
population reaches a mutation-selection balance  
 and the long term dynamics  \citep{Soderberg:2011,Lynch:2011} in
 which the 
mutators get fixed due to stochastic fluctuations have been 
studied. Here we have developed a
diffusion theory to find the time to fix mutators in a population of
size $N$ when the conversion rate $f$ is small in the presence of
epistatic interactions. Our results given in (\ref{TJfix}) predict
that the fixation time remains roughly  constant for small 
populations and increase exponentially fast beyond a critical
population size $N_c \sim  {\cal Q} f^{-1}$. Thus in the
limit $N \to \infty$, the time to fix mutators is also infinite which
is consistent with the result that an infinitely large population with
conversion rate $f \ll f_c$ exists in a mixed state. 
 The fixation time depends on the biological parameters $U, \lambda, s,
\alpha$ through the mutator fraction at mutation-selection balance. In  
regime I, since the mutator fraction  is independent of $\alpha$, 
this has the direct consequence that the
fixation time is also unaffected by $\alpha$ (refer
Fig.~\ref{TN1}) and therefore experiments aiming to detect epistasis
using mutators should not operate in regime I. From (\ref{Qss}), we
find that the crossover population size $N_c=(2 \delta U)^{-1}$ is
approximately $25$ for the parameters in Fig.~\ref{TN1} in agreement
with the simulation results shown in the inset. 
The fixation time in regime II depends on the epistasis exponent in 
a nontrivial manner. As discussed above, the mutator fraction in
regime II decreases with increasing 
$\alpha$ and therefore due to (\ref{TJfix}), it is expected that the
time to fix mutators increases and the crossover population size
decreases as $\alpha$ increases. These expectations are borne out by
the simulation results in Fig.~\ref{TN2}. We also note that since the
mutator fraction in regime II is higher than in regime 
I, the fixation time in regime II is shorter than that in regime
I. Although the diffusion theory developed here using simple
  arguments captures the
  essential features of the fixation time, a more sophisticated
  analysis based on a path-integral 
  formulation may be required for a better
  quantitative agreement with the numerical results \citep{Neher:2012}.

\subsection*{Evolution of mutation rates}

In this article, we
studied the dynamics of the evolutionary process by which the mutation
rate of a finite asexual population increases from $U$ to $\lambda U$
in time $T_1={\overline 
  T}$. This population can in turn produce mutators with mutation rate 
$\lambda_2 V$ at a rate $f_2$ which take an average time $T_2$ to fix
and so on.   
Thus the mutation rate of the population can keep increasing   in a
manner similar to Muller's ratchet \citep{Muller:1964} in which an
asexual population keeps accumulating deleterious mutations.  
This ratcheting process has been studied recently by
\citet{Soderberg:2011} and \citet{Lynch:2011} when epistatic
interactions are absent and mutator strength is weak (regime I). Here
we have focused on the dynamics of the first  click of the ratchet for
a more general set of parameters.  

When epistasis is absent,   
we can apply the results obtained here for the first
click of the ratchet to the succeeding clicks. Assuming that the conversion rate and
mutator strength is same for all clicks of the ratchet, the mutation rate at
the $n$th click of the ratchet 
will increase to $U_n= \lambda^n U$. Then using
(\ref{Qss}) and (\ref{Qws}) in the expression (\ref{TJfix}) for
fixation time, we find  
that the time elapsed between $(n-1)$th and $n$th click grows as 
 $e^{N \delta U_n}$ if $\delta U_{n-1} <
s/\lambda $ and $e^{N \sqrt{\delta U_n s}}$ otherwise, where $\delta
U_n=U_{n+1}-U_n$. Thus the time between
consecutive fixations increases but the rate of 
increase is slower in regime II than in regime I. 
Our analytical estimates above are consistent with the numerical results
of \citet{Soderberg:2011} who observed that the fixation time
increases with mutation rate $U_n$ and mutator strength $\lambda$. Our
results (\ref{Qss}) and (\ref{TJfix}) are also in agreement 
with those of \citet{Lynch:2011} who, 
using a diffusion theory and simulations, showed
that the mean time to fixation is roughly constant in $\delta U_n$ and
selective effects during the initial 
clicks characterised by $\delta U_n \ll 1/N$  but increases
exponentially fast for later clicks. 
On epistatic fitness landscapes, as the time between successive clicks
of the ratchet depends on the fittest class available at the end of a
click  \citep{Jain:2008b}, a quantitative understanding of the
successive fixation process requires more work and will be discussed 
elsewhere. 
Here we merely speculate that for $\alpha > 1/2$, every successive
fixation event will take longer to 
occur since the mutator fraction decreases with
increasing mutation rate and may result in fitness 
reduction to an extent that the population may not remain viable. 
On the other hand, as the mutator fraction at mutation-selection
balance increases with  mutation rate for $\alpha < 1/2$, the ratchet
is expected to click fast but the population may be able to tolerate
the errors as  the fitness penalty does not increase
commensurately. Interestingly in an experiment on bacterium {\it
    S. typhimurium} measuring the
  decrease in fitness as deleterious mutations accumulate, the mutation
rates was varied up to hundred times more than the wild type and the
fitness function of the form (\ref{fitness}) was found to be 
consistent with the epistasis exponent $\alpha \approx 1/2$ for
different mutation rates \citep{Maisnier-Patin:2005}.
Further experimental and theoretical
studies of the mutator dynamics on epistatic fitness landscapes are 
required to better understand the ratcheting process.

{\subsection*{Effect of beneficial mutations}

In the discussion so far,
  we have assumed that the backward conversion 
  rate $b$ at which a mutator is created from a nonmutator is
  zero. This assumption is motivated by experiments on {\it 
    E. coli} which indicate that $b/f \sim 10^{-3}$ 
  \citep{Ninio:1991}. Here we briefly discuss how a nonzero backward
  rate $b$ 
  affects our main results. An immediate consequence of the nonmutator
  creation is that the population remains in a mixed state, irrespective
  of its size. In the inset of Fig.~\ref{TN1}, the
  nonmutator fraction as a 
  function of time is  shown when the backward rate $b$ is nonzero.}
  We find that 
  the nonmutator frequency is high for about $4000$ generations after
  which the mutators take over. However as the nonmutators can be
  created at a rate $b \ll f$, the mutators cease to dominate after
  $6000$ generations and the process repeats. In
    Figs.~\ref{TN1} and \ref{TN2}, the  results
  of our numerical simulations
  measuring the average time at 
  which the nonmutator fraction becomes zero for the first time is
  compared with the fixation time for mutators when
  $b=0$ for a wide range of
    parameters and we find that the loss time for nonmutators is 
    mildly affected when the mutators are allowed to convert to
    nonmutators. However the 
  time during which  
  mutators dominate (data not shown) can not be obtained by 
  merely interchanging the nonmutator and mutator labels since the
  mutators by definition carry a higher mutation rate. A complete 
  analysis of this model is nontrivial and we hope to address it in a
  future study.

\begin{figure}
\begin{center}
\includegraphics[width=0.6 \linewidth,angle=270]{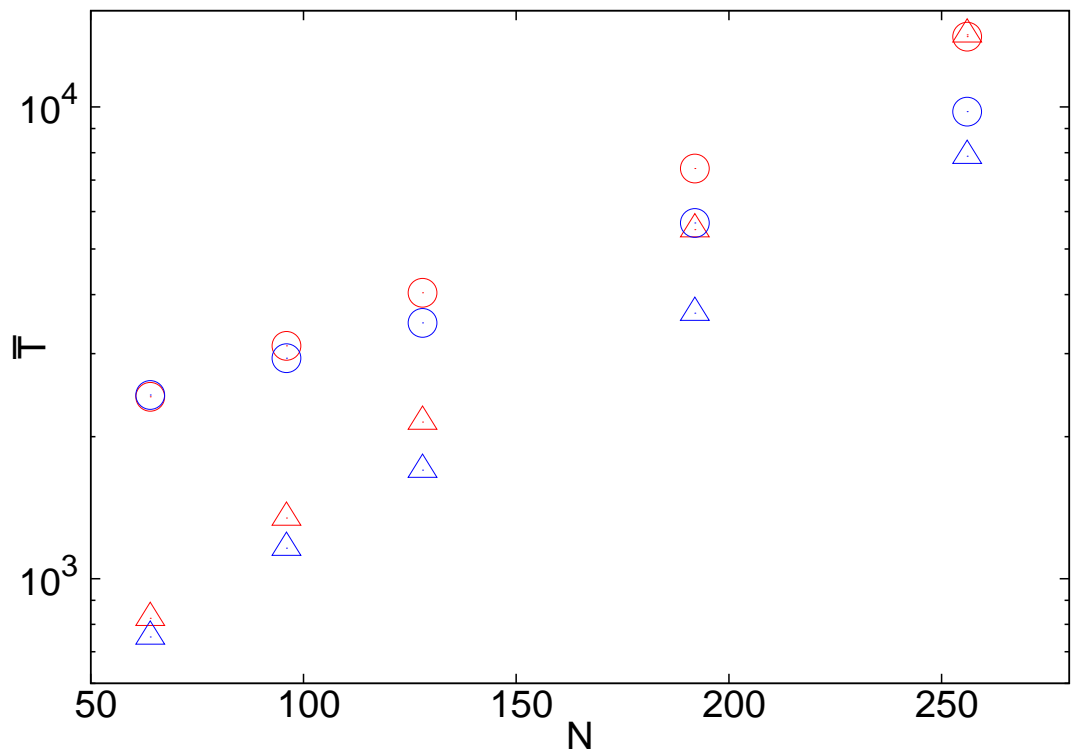}
\caption{Variation of fixation time with population size 
when rare beneficial mutations can also occur. The simulation data for
regime I ($f = 
    0.004, s = 0.2, U = 0.005, \lambda = 5, \alpha=1$) are shown by
    triangles and regime II ($f = 0.0005, s = 0.001, U = 0.005, \lambda =
    10, \alpha=1$) by circles. 
The blue points show the fixation time if a 
  deleterious mutation occurs at a rate $U (1-\epsilon)$ and a
  beneficial mutation at a rate $U \epsilon$ for $\epsilon=0.1$. 
  The red points show
the simulation data  in Figs.~\ref{TN1} and \ref{TN2} for $\alpha=1, b=0$ 
for comparison. }
\label{benf}
\end{center}
\end{figure}

Here we focused on the mutator fixation when the mutations are 
deleterious but theoretical studies
\citep{Andre:2006,Gerrish:2007,Wylie:2009} indicate that the mutators will
take over  during adaptation also. In this scenario, besides the continual
production of 
mutators and finite population size, an additional factor namely
availability of beneficial mutations is at work.  As a result, the
mutators are likely to get fixed in a  
time shorter than the one computed here. Indeed our numerical
  results shown in Fig.~\ref{benf} for a model in which beneficial 
  mutations can also occur are in agreement with this expectation.
  In a recent experiment by \citet{Raynes:2012}, the
number of generations for the 
mutators to take over an adapting population of {\it S. cerevisiae}
was measured. The fixation time was observed to increase with population
size but in very large populations, the mutators were not seen to
fix. 
The latter result is surprising since 
we have shown here 
that a finite asexual population will fix mutators even in the absence of
beneficial mutations. As the fixation time is 
large for larger populations, a longer run of the said experiment 
may shed some light on the dynamics of mutator fixation
during adaptation.

%============================================================================
%APPENDIX
%============================================================================

\section*{Appendix A: Infinite population on sharp peak fitness landscape}

When the exponent $\alpha \to 0$ in (\ref{fitness}), we get a sharp
peak fitness landscape defined as 
\be
F(k)= -s (1-\delta_{k,0})
\label{spl}
\ee
The conversion rate at which the transition occurs depends on the
nature of the pure mutator state. If the mutator
population is in the localised state, (\ref{fcdefn}) holds. But if  
the mutator population is uniformly distributed with an average
population fitness ${\cal {\widetilde W^*}}=-s$, we have 
\be
f_c^*=s-U ~,~ U < s
\label{fcstar}
\ee
Moreover, using (\ref{spl}) in (\ref{Pkss}) and (\ref{Qkss}), we
obtain 
\bea
{\cal P}(k) &=& {\cal P}(0) ~\left (\frac{U}{s} \right)^k ~,~ U < s
\label{Pkspl} \\
{\cal Q}(k) &=& \frac{V}{V-U+s-f} {\cal Q}(k-1)+\frac{f}{V-U+s
  (1-\delta_{k,0}) -f}
{\cal P}(k) 
\eea
On summing over $k$ from zero to infinity on both sides of the above
equations and using ${\cal P}+{\cal Q}=1$, we get 
\be
{\cal P}= \left(1-\frac{f}{f_c} \right)~\left(1-\frac{f}{f_c^*} \right) 
\ee
which shows that the total nonmutator fraction vanishes at both $f_c$
and $f_c^*$ in accordance with the above discussion and the pure
  mutator state is the only solution for $f > f_c, f_c^*$.

%AAAAAAAAAAAAAAAAAAAAAAAAAAAAAAAAAAAAAAAAAAAAAAAAAAAAAAAAAAAAAAAAAAAAAAAAAAA

\section*{Appendix B: Evaluation of product  (\ref{prod})}

The product (\ref{prod}) can be rewritten as 
\bea
G(k)
&=& \left( \frac{V}{\delta U} \right)^k {\rm exp} \left[- \sum_{p=0}^k
  \ln \left(1+ \frac{s p^\alpha}{\delta U} \right) \right] \\
&\approx& \left( \frac{V}{\delta U} \right)^k {\rm exp} \left[-
  \int_0^k dp~
  \ln \left(1+ \frac{s p^\alpha}{\delta U} \right) \right] \\
&=& \left( \frac{V}{\delta U} \right)^k {\rm exp} \left[-\left(\frac{\delta
    U}{s} \right)^{\frac{1}{\alpha}} \int_0^{k \left(\frac{s}{\delta
    U}\right)^{\frac{1}{\alpha}}} dz~\ln (1+ z^\alpha)
    \right]  
\label{prodA}
\eea
where we have approximated the sum by an integral. 
The function $G(k)$ is nonmonotonic in $k$ and
has a maximum at $k=k^*$ given by 
\be
k^* \approx \left( \frac{U}{s} \right)^{\frac{1}{\alpha}}
\label{kstar}
\ee
This can be seen by considering the ratio $G(k+1)/G(k)$ which crosses
unity at $k=k^*$. As the population fraction ${\cal P}_0(k)$ given by 
(\ref{Pksoln}) also has a maximum at $k^*$, the dominant contribution
to the sums in (\ref{Qcont}) comes when the argument  
$k \sim {\cal O}(k^*)$. As a result, the upper limit of the integral
on the RHS of  
(\ref{prodA}) scales as
\be
k \left(\frac{s}{\delta U}\right)^{\frac{1}{\alpha}} \sim \left(
\frac{U}{\delta U} \right)^\frac{1}{\alpha} \ll 1
\ee
if the mutator strength $\lambda \gg 1$. This allows us to write 
\bea
G(k) &\approx& \left( \frac{V}{\delta U} \right)^k {\rm exp}
\left[-\left(\frac{\delta 
    U}{s} \right)^{\frac{1}{\alpha}} \int_0^{k \left(\frac{s}{\delta
    U}\right)^{\frac{1}{\alpha}}} dz~z^\alpha
    \right] \\
&=&  \left( \frac{V}{\delta U} \right)^k {\rm exp} \left[-
  \frac{s}{\delta U} \frac{k^{\alpha+1}}{\alpha+1}\right]  
  \label{prodestim}
\eea

%AAAAAAAAAAAAAAAAAAAAAAAAAAAAAAAAAAAAAAAAAAAAAAAAAAAAAAAAAAAAAAAAAAAAAAAAAAA

\section*{Appendix C: Gaussian approximation in regime II}

To find the mutator fraction in regime II, we first note that 
both ${\cal P}_0(k)$ and $G(k)$ have 
a maximum at $k^*$ given by (\ref{kstar}). 
 Using the Stirling's formula $k! \approx \sqrt{2 \pi
  k} (k/e)^k$ in (\ref{Pksoln}), we approximate the fraction ${\cal
   P}_0(k)$ by a Gaussian centred about $k^*$ and obtain
\be
{\cal P}_0(k) \approx \sqrt{\frac{\alpha}{2 \pi k^*}}~e^{-\frac{\alpha
    (k-k^*)^2}{2 k^*}}
    \label{P0k}
\ee
which is normalised to unity for $k^* \gg 1$. The product (\ref{prodestim})
can also be approximated in a similar fashion and we have 
\bea
\ln \left(\frac{G(k)}{G(k^*)} \right) &\approx& 
 (k-k^*) \ln \left(\frac{V}{\delta U}
\right) -\frac{s}{\delta U} \frac{k^{\alpha+1}-{k^*}^{\alpha+1}}{\alpha+1} \\
&=&   (k-k^*) \ln \left(\frac{V}{\delta U}
\right) -\frac{s}{\delta U} \frac{{k^*}^{\alpha+1}}{\alpha+1} \left[
  \left(1+\frac{k-k^*}{k^*} \right)^{\alpha+1}-1\right]\\
&\approx& \left[\ln \left(\frac{V}{\delta U} \right) -\frac{U}{\delta U}\right]
(k-k^*) - \frac{\alpha U}{2 k^* \delta U}(k-k^*)^2
\eea
For strong mutators, we thus obtain
\be
G(k) \approx e^{- \frac{\alpha U (k-k^*)^2}{2 k^* \delta
    U}} G(k^*)
 \label{Gk}
\ee
Note that the width of the above Gaussian approximating the product $G(k)$
is about $\lambda$ times larger than that of the Gaussian in (\ref{P0k}). 

%CCCCCCCCCCCCCCCCCCCCCCCCCCCCCCCCCCCCCCCCCCCCCCCCCCCCCCCCCCCCCCCC

\section*{Appendix D: Calculation of fixation time (\ref{TJfix})}

Using the drift coefficient (\ref{close}) and diffusion coefficient
(\ref{D2}) in the expressions (\ref{Tdefn}) and (\ref{psi}), 
we get
\be
{\overline T}=2 N \int_0^{{\cal P}} dy ~\psi(y) \int_y^1 \frac{dx}{x (1-x)
  \psi(x)} 
  \label{Tapp}
\ee
where
\bea
\psi(x) &=& e^{-\frac{2 N x}{{\cal Q}_1}} (1-x)^{-2 N f} 
\eea
and ${\cal P} = 1-f {\cal Q}_1$. Thus we have 
\bea
{\overline T} &=& 2 N \int_0^{{\cal P}} dy ~e^{-\frac{2 N
    y}{{\cal Q}_1}} (1-y)^{-2 N f}~ \int_y^1 dx~ \frac{ e^{\frac{2 N
      x}{{\cal Q}_1}} (1-x)^{2 N f}  }{x (1-x)}  
\label{Tfull}
\eea

We first evaluate the integrals in the above equation for $2 N f <
1$ by writing ${\overline T}=I_1+I_2$ where 
\bea
I_1 &=& 2 N \int_0^{{\cal P}} dy ~e^{-\frac{2 N
    y}{{\cal Q}_1}} (1-y)^{-2 N f}~ \int_y^1 dx~ e^{\frac{2 N
      x}{{\cal Q}_1}} (1-x)^{2 N f-1} \label{I1} \\
I_2 &=& 2 N \int_0^{{\cal P}} dy ~e^{-\frac{2 N
    y}{{\cal Q}_1}} (1-y)^{-2 N f}~ \int_y^1 dx~ e^{\frac{2 N
      x}{{\cal Q}_1}} (1-x)^{2 N f} x^{-1} \label{I2}
\eea
Since the factor $(1-x)^{2 N f-1}$ in the inner
integral on the RHS of (\ref{I1}) diverges at unity for $2 N f < 1$,
the dominant contribution to the integral $I_1$ comes due to this
factor:
\bea
I_1 &\approx& 2 N \int_0^{{\cal P}} dy ~e^{-\frac{2 N
    y}{{\cal Q}_1}} (1-y)^{-2 N f}~ \int_y^1 dx~ e^{\frac{2 N
      }{{\cal Q}_1}} (1-x)^{2 N f-1} \\
&=&  \frac{{\cal Q}_1}{2 N f} \left(e^{\frac{2 N}{{\cal Q}_1}}-e^{2 N
  f} \right) 
\eea
If $N \ll {\cal Q}_1/2$, the above integral gives the fixation time  
to be independent of $N$ {\it i.e.}
\bea
I_1 \approx \frac{1}{f}-{\cal Q}_1 =\frac{{\cal P}}{f} ~,~ N
\ll {\cal Q}_1/2 
\label{smallNI1}
\eea
On the other hand, for small $2 N f$ and $2 N/{\cal Q}_1$, the integral $I_2$ 
may be approximated as:
\bea
I_2 &\approx& 2 N \int_0^{{\cal P}} dy \int_y^1 dx~x^{-1} \\
&\approx & 2 N {\cal P} ~,~N \ll {\cal Q}_1/2 
\label{smallNI2}
\eea
Since $2 N f < 1$, the behavior of the fixation time for $N \ll {\cal
  Q}_1/2$ is dominated by (\ref{smallNI1}). Thus the diffusion
theory predicts that the fixation time
for small populations is a constant in $N$. We remark that the
$N$-independent behavior of fixation time will not be obtained if
instead of (\ref{D2}), Feller diffusion $D_2=P/N$ is assumed (see, for
e.g., \citet{Jain:2008b}).    

For large populations with $2 N f, 2 N/{\cal Q}_1 \gg 1$, a rough
estimate for the fixation time may be 
obtained using (\ref{Tfull}). Approximating $(1-x)^{2 Nf} \approx
e^{-2 N f x}$, we get 
\bea
{\overline T} &\approx& 2 N \int_0^{{\cal P}} dy ~e^{-\frac{2 N
    y}{{\cal Q}_1}} e^{2 N f y}~ \int_y^1 dx~ e^{\frac{2 N
      x}{{\cal Q}_1}} e^{-2 N f x} \\
&\approx& \frac{{\cal Q}^2_1}{2 N {\cal P}^2} ~e^{\frac{2 N {\cal P}}{{\cal Q}_1}}
\eea
Thus for large $N$, the fixation time increases exponentially in $N$.

%============================================================================

\subsection*{Acknowledgements} A.N. acknowledges the kind hospitality
of JNCASR during his visits.

%============================================================================

%============================================================================
%FIGURES
%============================================================================

\end{document}